\pdfoutput=1
\documentclass[a4paper,fleqn,usenatbib]{mnras}
\usepackage{amsmath}
\usepackage{tablefootnote}
\usepackage{natbib}
\usepackage{bm}
\usepackage{verbatim}
\usepackage{graphicx}
\usepackage{mathtools, cuted}
\usepackage[T1]{fontenc}
\usepackage{lmodern}
\usepackage{amssymb}
\usepackage{flushend}
\makeatletter
\patchcmd{\NAT@citex}
  {\@citea\NAT@hyper@{%
     \NAT@nmfmt{\NAT@nm}%
     \hyper@natlinkbreak{\NAT@aysep\NAT@spacechar}{\@citeb\@extra@b@citeb}%
     \NAT@date}}
  {\@citea\NAT@nmfmt{\NAT@nm}%
   \NAT@aysep\NAT@spacechar\NAT@hyper@{\NAT@date}}{}{}

\patchcmd{\NAT@citex}
  {\@citea\NAT@hyper@{%
     \NAT@nmfmt{\NAT@nm}%
     \hyper@natlinkbreak{\NAT@spacechar\NAT@@open\if*#1*\else#1\NAT@spacechar\fi}%
       {\@citeb\@extra@b@citeb}%
     \NAT@date}}
  {\@citea\NAT@nmfmt{\NAT@nm}%
   \NAT@spacechar\NAT@@open\if*#1*\else#1\NAT@spacechar\fi\NAT@hyper@{\NAT@date}}
  {}{}

\makeatother

\defcitealias{caldwell2017dust}{CHK17}
\defcitealias{goldreich1995toward}{GS95}

\title[Ratio of E to B mode power for dust galactic foreground]{Can the observed E/B ratio for dust galactic foreground be explained by sub-Alfv\'enic turbulence?}

\author[Kandel, Lazarian \& Pogosyan]{
D. Kandel,$^{1}$
A. Lazarian$^{2}$
and D. Pogosyan$^{1}$
\\
$^{1}$Physics Department, University of Alberta, Edmonton, T6G 2E1, Canada\\
$^{2}$Department of Astronomy, University of Wisconsin, 475 North Charter Street, Madison, WI 53706, USA\\
}

\date{Accepted 2017 August 08 . Received 2017 August 08 ; in original form 2017 July 12}

\pubyear{2017}

\begin{document}
\label{firstpage}
\pagerange{\pageref{firstpage}--\pageref{lastpage}}
\maketitle

\begin{abstract}
Recent Planck observations of dust polarization in the Galaxy have revealed that the  power in $E$ mode is twice that in $B$ mode.  Caldwell et al. have formulated a theoretical model in the context of magnetohydrodynamic (MHD) turbulence and found it problematic to account for this result. In particular, they concluded that there is a very narrow range of theoretical parameters that could account for the observation. This poses a problem of whether the accepted description of MHD turbulence can apply to the interstellar medium. We revisit the problem and demonstrate that MHD turbulence corresponding to the high galactic latitudes range of Alfv\'en Mach numbers, i.e.  $M_A\lesssim 0.5$, can successfully explain the available results for the $E$ to $B$ mode ratio. 
\end{abstract}
\section{Introduction}

The dominant polarized foreground of the cosmic microwave background (CMB) above 100 GHz comes from thermal emission by aligned dust grains. These grains get aligned with longest dimension perpendicular to magnetic field (see \citealt{lazarian2007tracing} for a review) and, as a result, the polarization traces interstellar magnetic fields weighted by the dust density. These magnetic and density fields are expected to be turbulent (see \citealt{armstrong1995electron}; \citealt{chepurnov2010extending}; see also \citealt{2007ARA&A..45..565M} for a review), and therefore dust polarization measurement can be an important probe to turbulence in the interstellar medium (ISM). Moreover, dust polarization interferes with the CMB polarization measurements, thus the foreground contribution of dust polarization should be properly accounted for successful CMB polarization studies. 

A linear dust polarization map can be decomposed into two rotationally invariant modes, $E$ mode and the $B$ mode. For a randomly oriented polarization map, the two modes are expected to have equal power. The recent Planck measurement \citep{2016A&A...586A.133P} of dust polarization at 353 GHz, however, has shown different results: first, the ratio of power in E mode is twice to that in $B$ mode, and secondly, there is a positive temperature $E$-mode ($TE$) cross-correlation. \citet[hereafter \citetalias{caldwell2017dust}]{caldwell2017dust} have tried to explain the observed $EE/BB$ ratio and TE correlation in the context of magnetohydrodynamical turbulence using the description of turbulence in the observer's frame (see \citealt{lazarian2012statistical}). Their main conclusion is that at large scale turbulence could be unimportant, as there is a very narrow range of parameters in the theoretical model that could possibly mimic the Planck result. This result perturbed the community as it questions the generally accepted notion of turbulent origin of dust-polarization fluctuations. 

In this Letter we revisit the problem and argue that with realistic ISM conditions, the range of theoretical parameters that mimics the Planck result does not contradict to what we know about the ISM at high galactic latitudes. In fact, we show that the Planck result can be explained in the context of MHD turbulence if the Alfv\'en Mach number in the high Galactic latitude is less than $0.5$. The latter number is in agreement with the existing expectations (see \citealt{2016A&ARv..24....4B}). 

The structure of our Letter is the following. In Sec. \ref{mhdmodes}, we briefly review three fundamental modes, Alfv\'en, fast and slow, of magnetohydrodynamic turbulence, and write down expressions of fluctuations of magnetic field and density induced by these modes. In Sec. \ref{model}, we present the model we adopt in the Letter, and write down relevant expressions for the power of $E$ and $B$ mode.  We present our results for $E$ and $B$ power ratio for three fundamental MHD modes in Sec. \ref{sec:results}, and discuss these results in Sec. \ref{sec:disc}.

\section{Mode description of MHD turbulence}\label{mhdmodes}
The work of \citet[hereafter \citetalias{goldreich1995toward}]{goldreich1995toward} has made substantial progress in the theory of incompressible MHD turbulence. The \citetalias{goldreich1995toward} model predicts a Kolmogorov velocity spectrum and scale-dependent anisotropy, and these predictions have been confirmed numerically (\citealt{cho2000anisotropy}; \citealt{maron2001simulations}), and are in good agreement with observations (see \citealt{vishniac2003problems}). A important property of incompressible MHD turbulence is  that fluid motions perpendicular to magnetic field are identical to hydrodynamic motions, thus providing a physical insight as to why in some respect MHD turbulence and hydrodynamic turbulence are similar, while in other respect they are different.

MHD turbulence is in general compressible. The description of incompressible MHD turbulence was extended to account for compressibility of turbulent media in \citet{cho2002compressible, cho2003compressible} by decomposing motions into basic MHD modes (Alfv\'en, slow and fast). The Alfv\'enic and slow modes keep the scaling and anisotropy of the incompressible MHD, while fast modes shows different scaling and exhibits isotropy in power spectrum.

The properties of MHD turbulence depend on the degree of magnetization, which can be characterized by the Alfv\'en Mach number $M_A = V_L/a$, where $V_L$ is the injection velocity at the scale $L$ and $a$ is the Alfv\'en velocity. For super-Alfv\'enic turbulence, i.e. $M_A\gg 1$, magnetic forces should not be important in the vicinity of injection scale, thus corresponding to the case of marginally perturbed magnetic field. For sub-Alfv\'enic turbulence, i.e. $M_A<1$, magnetic forces are important, and turbulence statistics are highly anisotropic. Another important parameter is the plasma $\beta$ ($\equiv P_{\text{gas}}/P_{\text{mag}}$), which characterizes compressibility of a gas cloud. Formally, $\beta\rightarrow\infty$ denotes incompressible regime.

Here, we briefly describe statistical properties of the fluctuations of magnetic and density fields induced by motions in three MHD modes. These statistical properties are determined by the allowed plasma displacements in each mode. To facilitate comparison with \citetalias{caldwell2017dust}, we use their notations. The line of sight (LOS) is assumed to be along the $z-$axis, and the mean field $\bm{H}_0=H_0(\sin\theta, 0, \cos\theta)$ is assumed to be aligned in the $x-z$ plane making an angle $\theta$ with the LOS. We consider perturbations with two-dimensional wavevector $\bm{K}=K(\cos\psi, \sin\psi, 0)$ in the $x-y$ plane of the sky, as observations effectively give the two dimensional sky maps. The angle $\alpha$ between wavevector and magnetic field is then $\cos\alpha=\sin\theta\cos\psi$. 

The power spectrum of magnetic field perturbation is given by (see \citealt{cho2002compressible}; \citetalias{caldwell2017dust})
\begin{equation}\label{pihk}
P_{i,H}(k, \alpha)=P_i(k)F_i(\alpha)\left[h_i(\alpha)\right]^2~,
\end{equation}
where $i=\{\text{a, f, s}\}$, denotes Alfv\'en, fast and slow modes. The $h_i(\alpha)$ describes anisotropic tensor structure of each modes as determined by the allowed displacement in a plasma (see \citealt{lazarian2012statistical}). Specifically,
\begin{equation}
h_\text{a}(\alpha)=\frac{1}{a}~,
\end{equation}
where $a\equiv H_0/\sqrt{4\pi\rho_0}$ is the Alfv\'en speed. Additionally, 
\begin{equation}\label{hf}
h_\text{f}(\alpha)=\frac{k}{\omega}\frac{\sin\alpha}{\left(\zeta_f^2\cos^2\alpha+\sin^2\alpha\right)^{1/2}}~,
\end{equation}
and 
\begin{equation}\label{hs}
h_\text{s}(\alpha)=\frac{k}{\omega}\frac{\zeta_s\sin\alpha}{\left(\cos^2\alpha+\zeta_s^2\sin^2\alpha\right)^{1/2}}~.
\end{equation}
In addition, the fast and slow modes satisfy the following dispersion relation
\begin{equation}
\left(\frac{\omega}{k}\right)^2=\frac{a^2}{2}(1+\beta/2)\left[1\pm\left(1-\frac{2\beta\cos^2\alpha}{(1+\beta/2)^2}\right)^{1/2}\right]~,
\end{equation}
where plus sign is for fast mode and minus for slow mode. 
In equations \eqref{hf} and \eqref{hs},  
\begin{align}
\zeta_\text{f}&=\frac{1-\sqrt{D}+\beta/2}{1+\sqrt{D}-\beta/2}\tan^2\alpha~,\nonumber\\
\zeta_\text{s}&=\frac{1-\sqrt{D}-\beta/2}{1+\sqrt{D}+\beta/2}\cot^2\alpha~,\nonumber\\
D&=(1+\beta/2)^2-2\beta\cos^2\alpha~.
\end{align}
In equation \eqref{pihk}, $P_i(k)F_i(\alpha)$ describes anisotropic power, which contains factorised scale dependent part and an angle dependent part. Since the power spectrum of fast mode is isotropic
\begin{equation}\label{fastpowereq}
F_\text{f}(\alpha)=1~,
\end{equation}
while the anisotropic part of the power spectrum applicable for Alfv\'en and slow modes is given by (\citetalias{goldreich1995toward})
\begin{equation}\label{anisgs}
F_{\text{a,s}}(\alpha)=\exp\left[-\frac{M_A^{-4/3}\cos^2\alpha}{(\sin^2\alpha)^{2/3}}\right]~.
\end{equation}
Note that \citetalias{caldwell2017dust} use a toy model 
\begin{equation}\label{toygs}
F_{\text{a,s}}(\alpha)=(\sin^2\alpha)^{-\lambda}~,\qquad \lambda<0
\end{equation}
which mimics the fact that fluctuations along the direction of magnetic field are suppressed.

Within our model, Alfv\'en modes cannot induce any density fluctuations, the only contribution to density fluctuations comes from fast and slow compressible modes. At linear level, applying continuity equation together with frozen-in condition for magnetic field in plasma, gives density perturbations produced by fast and slow modes as
\begin{equation}\label{deltanr}
\frac{\delta n_i}{n_0}=g_i(\alpha)h_i(\alpha)|\bm{v}|=g_i(\alpha)\frac{|\delta \bm{H}|}{H_0}~,
\end{equation}
where 
\begin{equation}
g_\text{f}(\alpha)=\frac{\zeta_f\cos^2\alpha+\sin^2\alpha}{\sin\alpha}~,
\end{equation}
and 
\begin{equation}
g_\text{s}(\alpha)=\frac{\cos^2\alpha+\zeta_s\sin^2\alpha}{\zeta_s\sin\alpha}~.
\end{equation}
Thus, the power spectrum of density fluctuations is
\begin{equation}
P_\rho(k, \alpha)=P_i(k)F_i(\alpha)\left[g_i(\alpha)h_i(\alpha)\right]^2~,
\end{equation}
We remark that equation~\eqref{deltanr} should not be over-interpreted 
to mean that gas density and magnetic field fluctuations are not independent
degrees of freedom. Indeed,  equation~\eqref{deltanr} is written in Fourier -- 
frequency $\mathbf{k},\omega$ domain, and has a structure 
$\omega \delta n(\mathbf{k},\omega) \propto \omega \delta H(\mathbf{k},\omega)$.
On the fixed time hypersurface, it
relates time derivatives of the gas density and the magnetic field, while there is still a free time independent
function that can be added in the relation between density and magnetic field themselves.
How correlated the density and magnetic field fluctuations 
at fixed time are in the developed turbulence is a subject for discussion.
In general, the correlation depends on the sonic Mach number, since it determines whether or not density clumps will develop \citep{burkhart2009density}.
Numerical studies show that for sub-sonic case, density and magnetic field are
weakly correlated, while correlation start to become important for super-sonic
turbulence (see \citealt{burkhart2009density}). 
Turbulence in the galaxy is expected to be sub-sonic, at least for the warm gas, and therefore, it is not unreasonable to consider the limit when density and magnetic field are not tightly correlated.

The model of MHD turbulence we presented has the following main parameters: $M_A$, and $\beta$. Anisotropy is highly sensitive to $M_A$ (see equation \ref{anisgs}). Although equation \eqref{toygs} also describes correct qualitative features of anisotropies, it is $M_A$ which is a physical parameter (formally $\lambda\gg 1$ corresponds to $M_A\ll 1$, both of which are highly anisotropic). Ultimately, observational data should be used to place limits in space of physical parameters, $M_A$ and $\beta$. Another consideration that needs to be made is the model of density fluctuations, and their correlation with magnetic field. 

\section{E and B modes induced by dust polarization}\label{model}
The two-dimensional projection of polarized emission is assumed to have a form of 
\begin{equation}\label{pemiss}
\epsilon_P=\epsilon_Q + i\epsilon_U=An_dH^{\gamma}(H_x+iH_y)^2~,
\end{equation}
where $\gamma$ is an exponent\footnote{Similar ansatz has been used for synchrotron emission, where $n_d$ plays the role of relativistic electrons (see \citealt{lazarian2012statistical}).}. For $\gamma=-2$, the polarized emission is independent of magnetic field strength. In equation \eqref{pemiss}, $n_d$ is dust density which we take to be proportional to gas density (see \citealt{lazarian2002grain}), and $A<0$ is a constant, and its value is taken to be negative so that the dust polarization is perpendicular to the direction of magnetic field.

For a given Fourier mode of wavevector $K$ transverse to the LOS in a box of radial width $\Delta r$, the $E$ and $B$ modes at the multipole $l = K \Delta r$
will have the form (\citetalias{caldwell2017dust})
\begin{align}\label{emain}
\tilde{E}=An_0H_0^{2+\gamma}\frac{\Delta r}{r^2}\big[-\sin 2\theta\frac{\sin\psi}{\sin\alpha}\frac{\delta H_\text{a}}{H_0}\nonumber\\
+\frac{\sin^2\theta\left[-2\sin^2\psi(1+\gamma\sin^2\alpha)+\gamma\sin^2\alpha\right]}{\sin\alpha}\frac{\delta\tilde{H}_p}{H_0}\nonumber\\
+\sin^2\theta\cos 2\psi\frac{\delta\tilde{n}_d}{n_0}\big]~,
\end{align}
\begin{align}\label{bmain}
\tilde{B}=An_0H_0^{2+\gamma}\frac{\Delta r}{r^2}\big[-\sin 2\theta\frac{\cos\psi}{\sin\alpha}\frac{\delta H_\text{a}}{H_0}\nonumber\\
-\frac{2\sin^2\theta\sin\psi\cos\psi(1+\gamma\sin^2\alpha)}{\sin\alpha}\frac{\delta\tilde{H}_c}{H_0}\nonumber\\
-\sin^2\theta\sin 2\psi\frac{\delta\tilde{n}_d}{n_0}\big]~,
\end{align}
where $\delta n_d/n_0$ is the fractional dust density perturbation, $\delta H_\text{a}/H_0$ and $\delta H_p/H_0$ are the fractional magnetic field perturbations due to Alfv\'en modes and compressible (fast and slow) modes.
In this Letter, we explore the case of uncorrelated dust density and magnetic field in comparison to the completely correlated prescription used in \citetalias{caldwell2017dust}. Quantitatively, we use equation~\ref{deltanr} to estimate the minimum level of amplitude that density fluctuations will develop in turbulence;
but we still treat density and magnetic field contributions to observables
to be uncorrelated. Our further results show that E/B ratio is not very sensitive
to the level of correlation between $\delta n_d$ and $\delta H$.

Using equations \eqref{emain} and \eqref{bmain}, one can easily show that the power of $E$ and $B$ mode for Alfv\'en wave is
\begin{equation}
\tilde E^2\propto \sin^22\theta\frac{\sin^2\psi}{\sin^2\alpha}P_H+\sin^4\theta\cos^22\psi P_\rho~,
\end{equation} 
\begin{equation}
\tilde B^2\propto \sin^22\theta\frac{\cos^2\psi}{\sin^2\alpha}P_H+\sin^4\theta\sin^22\psi P_\rho~,
\end{equation}
where $P_\rho\equiv \langle (\delta n/n_0)^2\rangle$ and $P_H\equiv\langle(\delta H/H_0)^2\rangle$ are power spectrum of fractional density and magnetic field perturbations, respectively.

Similarly, it can be shown that for compressible modes,
\begin{align}\label{sfm}
\tilde E^2\propto  \frac{\sin^4\theta\left[-2\sin^2\psi(1+\gamma\sin^2\alpha)
+\gamma\sin^2\alpha\right]^2}{\sin^2\alpha}P_H\nonumber\\
+\sin^4\theta\cos^22\psi P_\rho~,
\end{align}
and
\begin{equation}\label{sfmb}
\tilde B^2\propto \frac{\sin^4\theta\sin^22\psi(1+\gamma\sin^2\alpha)^2}{\sin^2\alpha}P_H+\sin^4\theta\sin^22\psi P_\rho~.
\end{equation}
The main quantity of interest for us is the ratio of power in $E$ mode to that in $B$ mode. This ratio has been obtained from the Planck survey at high galactic latitudes. Formally, we characterize this ratio as 
\begin{equation}
R=\frac{\int\mathrm{d}\Omega\tilde{E}^2}{\int\mathrm{d}\Omega\tilde{B}^2}~,
\end{equation}
where each integral represent three dimensional angular averaging.
\section{Results} \label{sec:results}
\begin{figure*}
\centering
\includegraphics[scale=0.4]{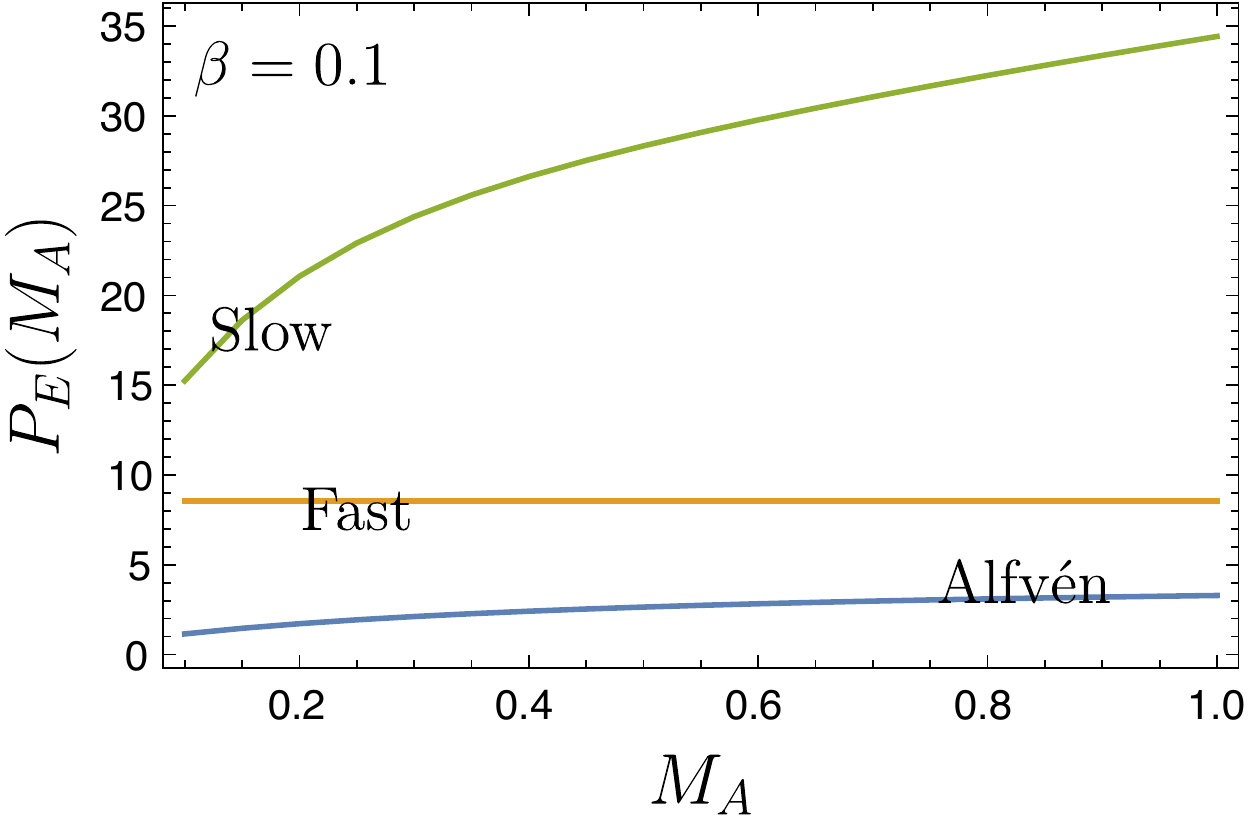}\hspace*{0.1cm}
\includegraphics[scale=0.4]{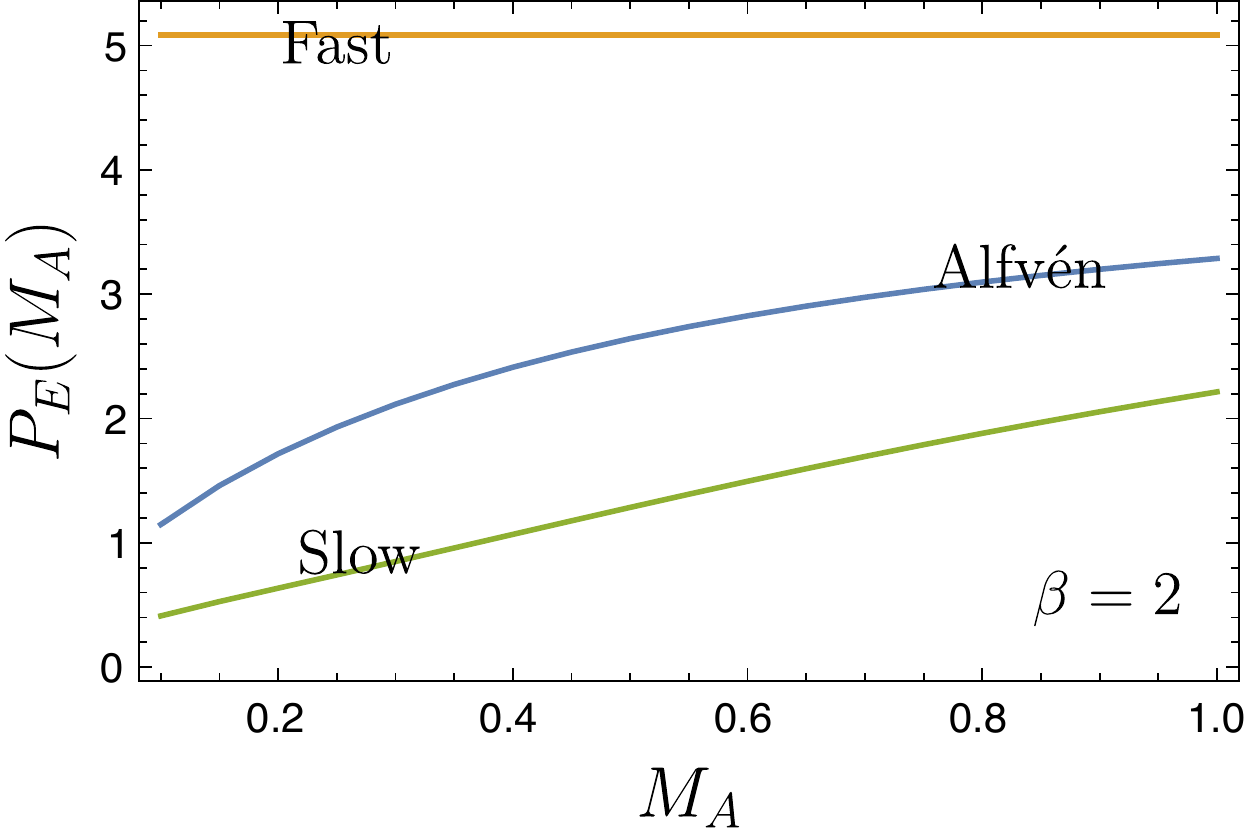}\hspace*{0.1cm}
\includegraphics[scale=0.4]{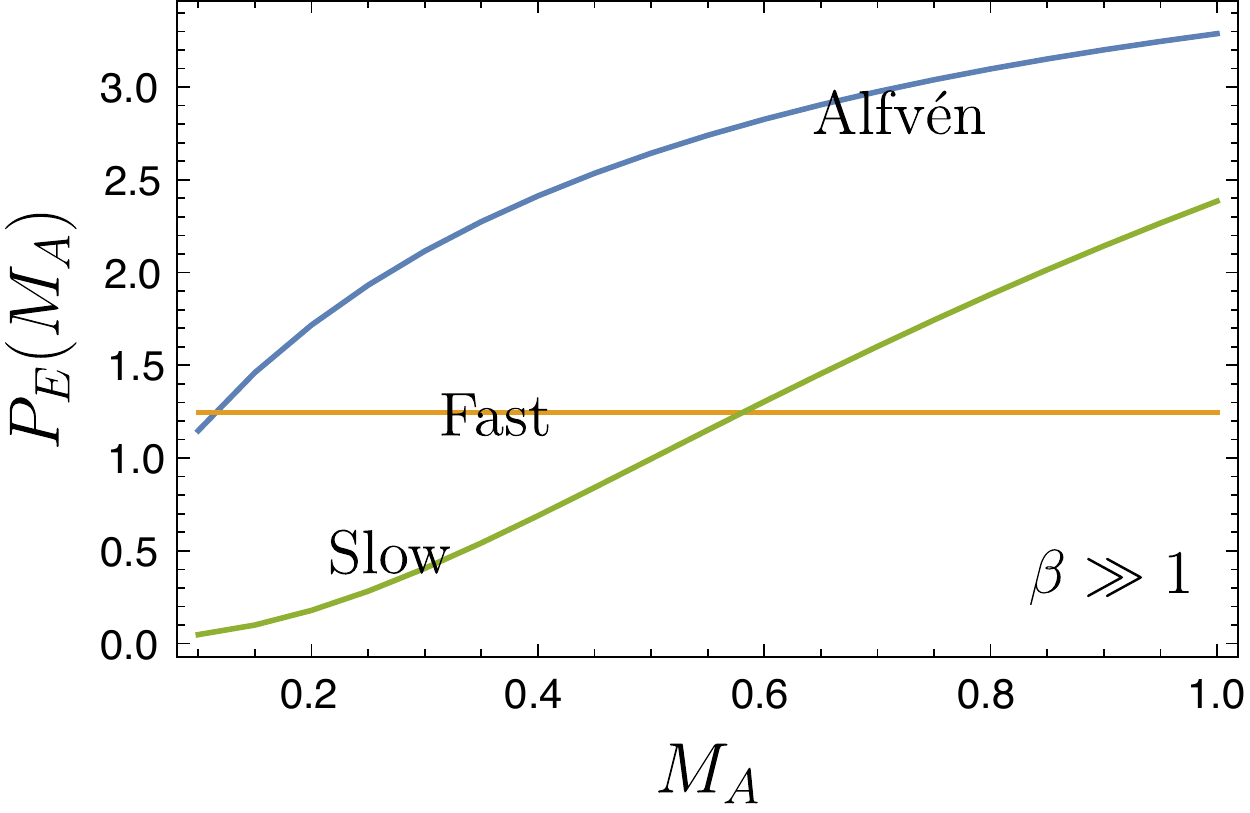}
\vspace*{0.1 cm}
\includegraphics[scale=0.4]{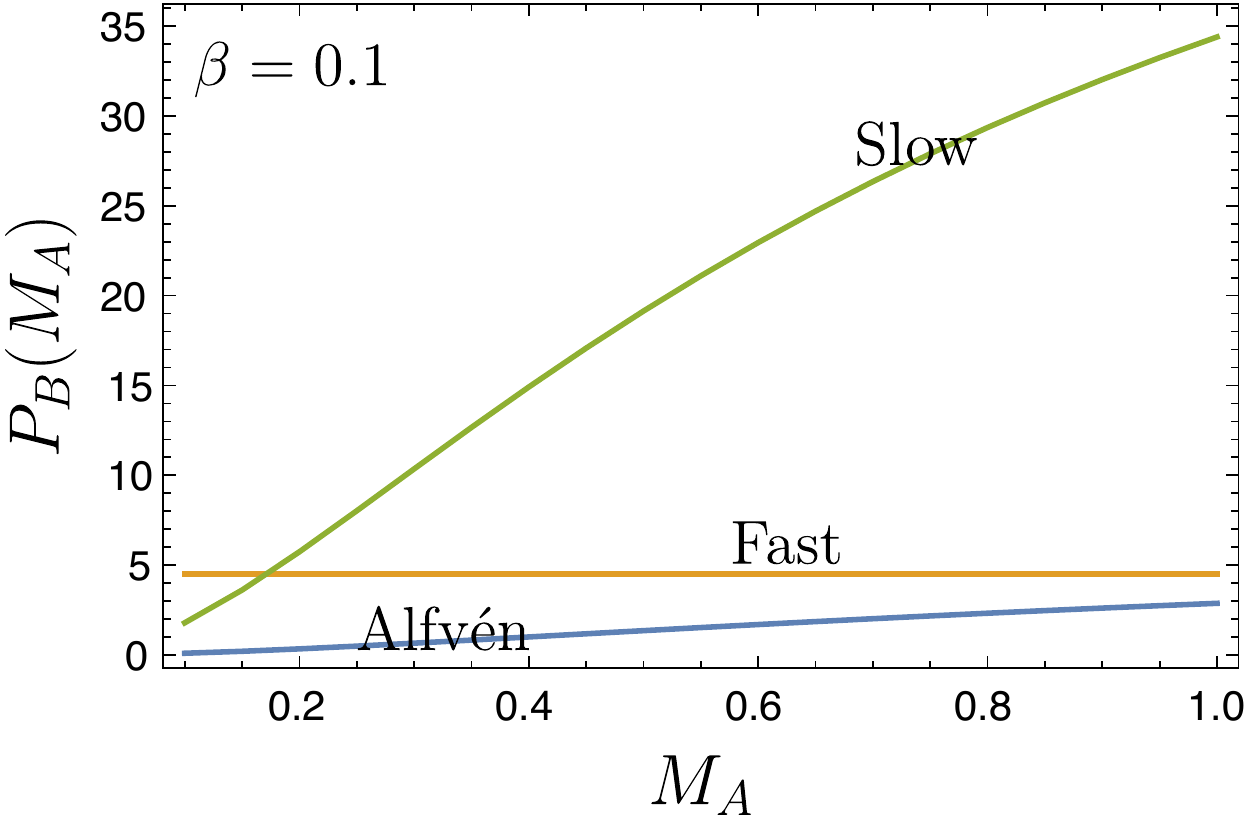}\hspace*{0.1cm}
\includegraphics[scale=0.4]{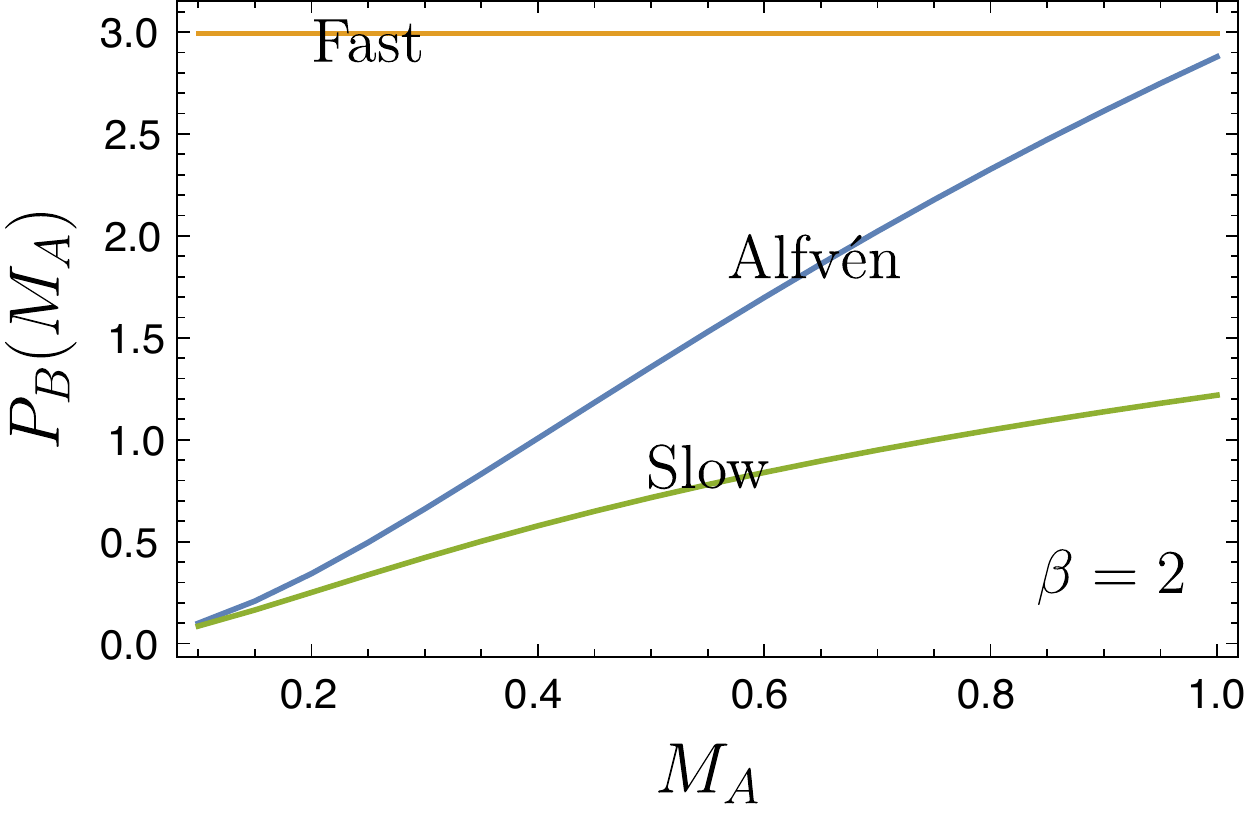}\hspace*{0.1cm}
\includegraphics[scale=0.4]{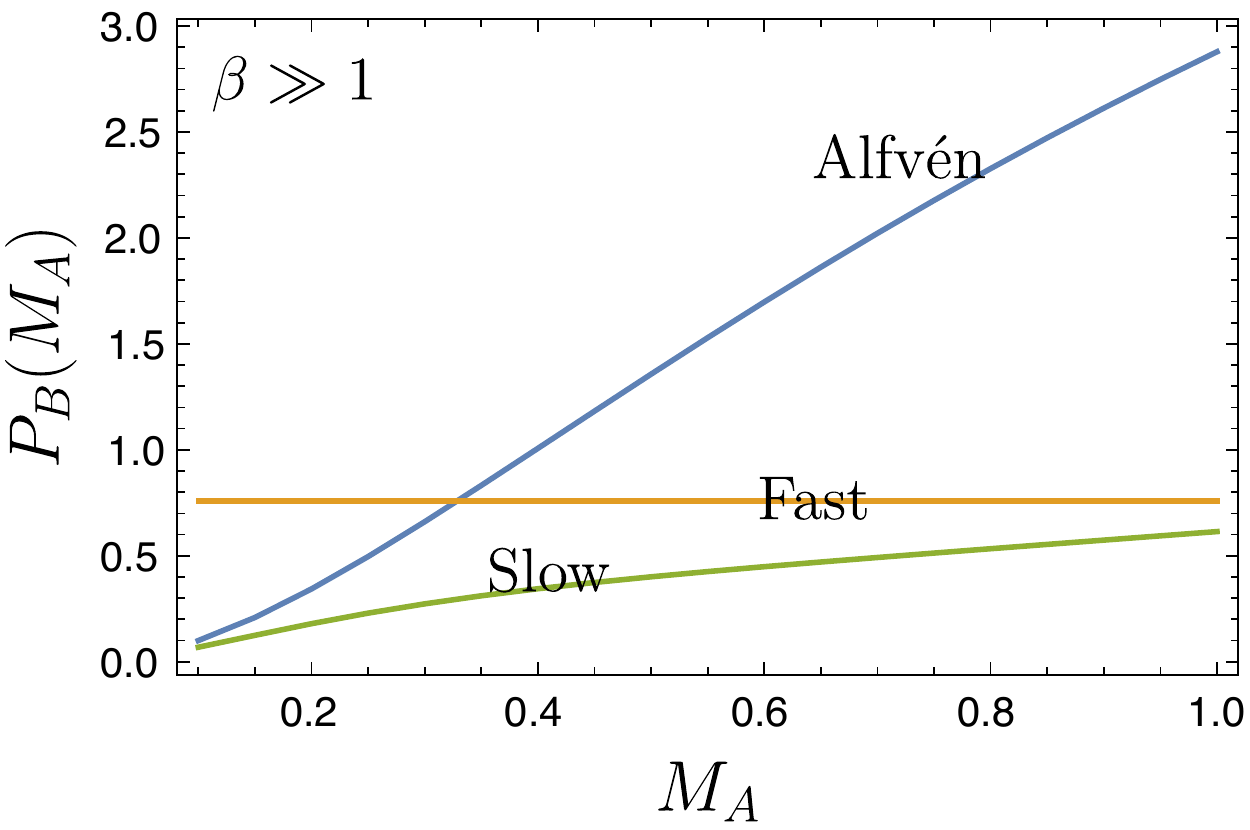}
\vspace*{0.1cm}
\includegraphics[scale=0.4]{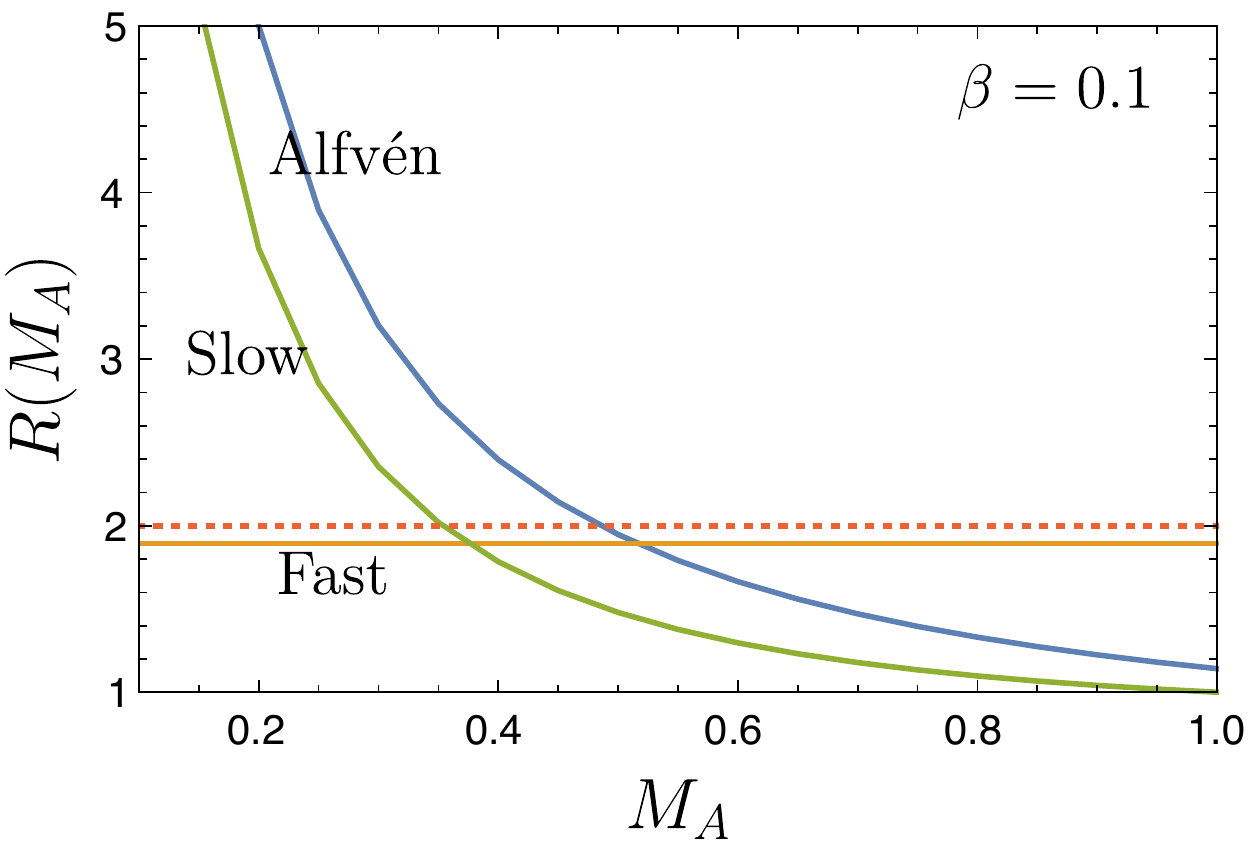}\hspace*{0.1cm}
\includegraphics[scale=0.4]{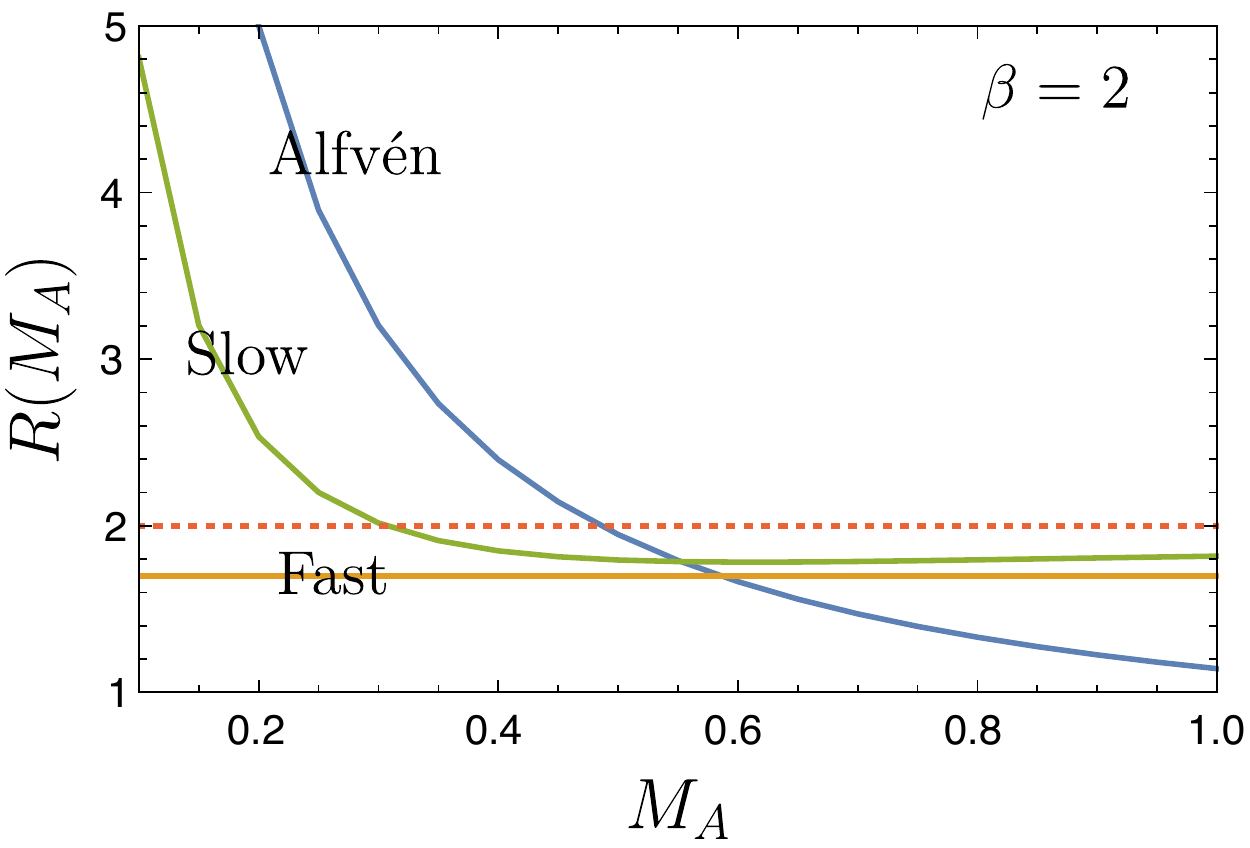}\hspace*{0.1cm}
\includegraphics[scale=0.4]{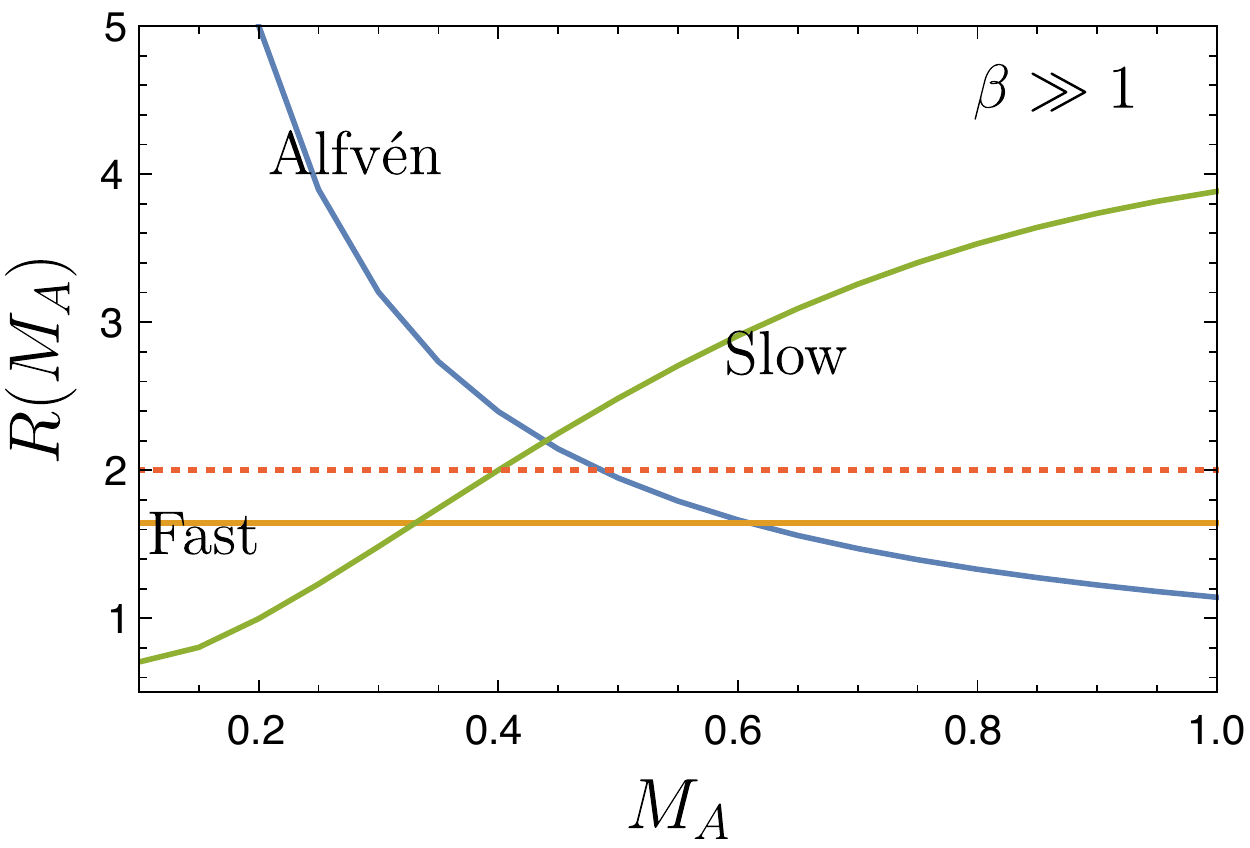}
\caption{Upper row: $E$ power amplitude for three modes at various $\beta$. Centre row: $B$ power amplitude for three modes at various $\beta$. In both these panels, $E$ and $B$ are in arbitrary units, and $\gamma=-2$. Lower row: ratio of $E$ to $B$ power for three modes at various $\beta$. The dotted line represents the observed $E$ to $B$ power ratio of 2.}
\label{threemachplots}
\end{figure*}

\begin{figure*}
\centering
\includegraphics[scale=0.4]{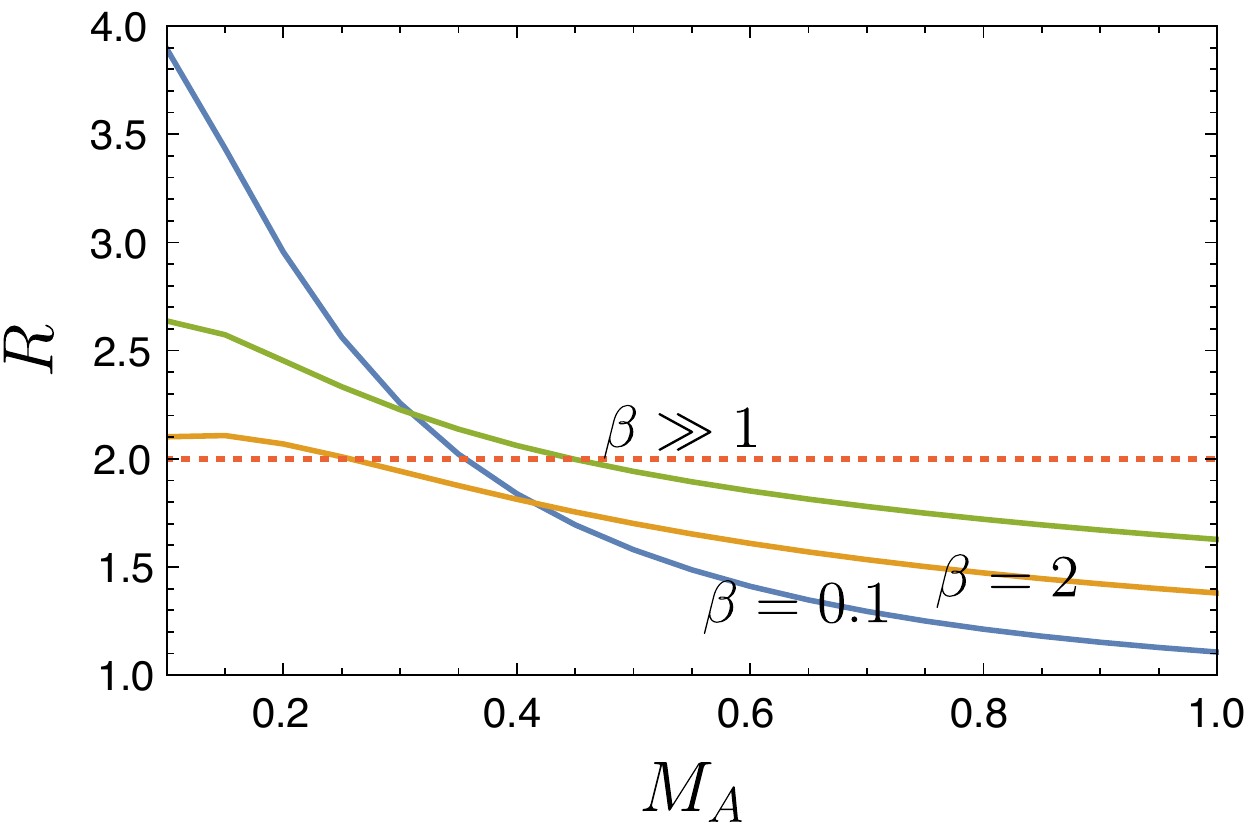}\hspace*{0.2cm}
\includegraphics[scale=0.4]{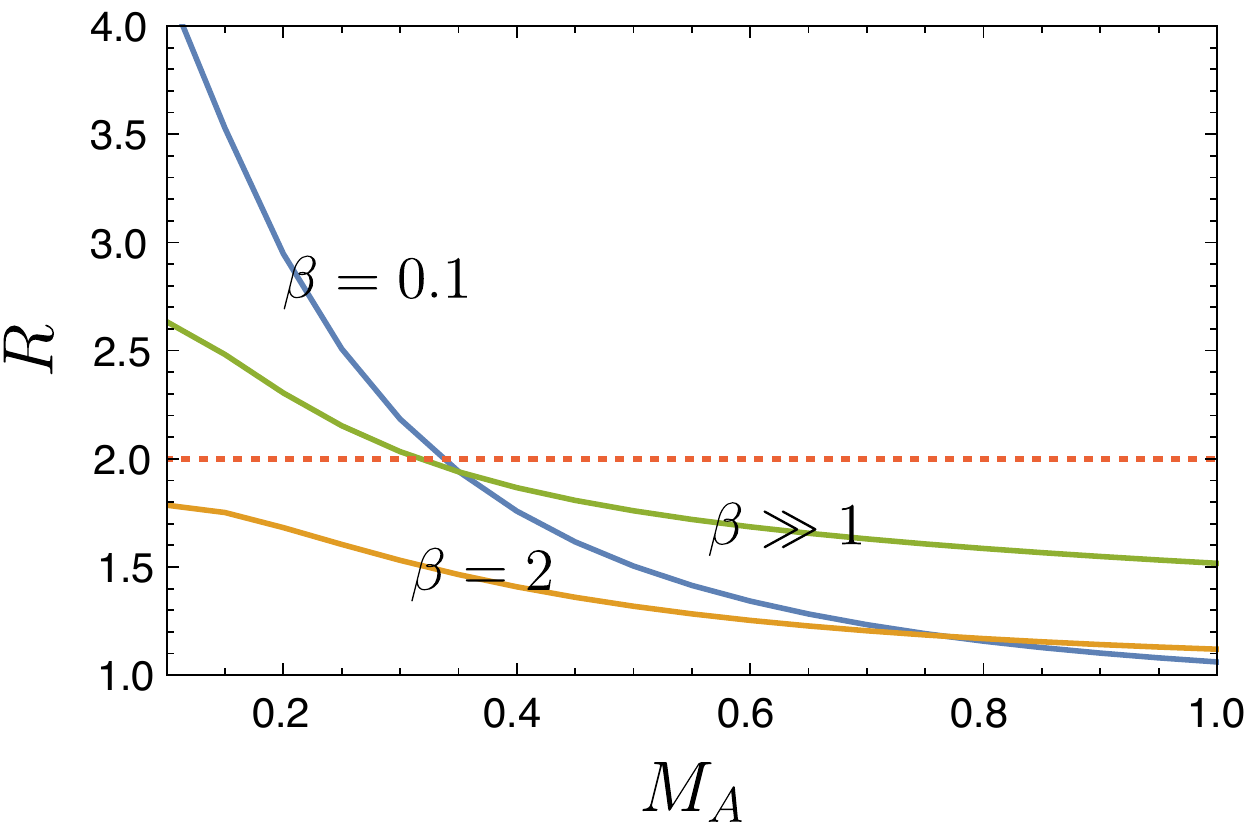}
\includegraphics[scale=0.4]{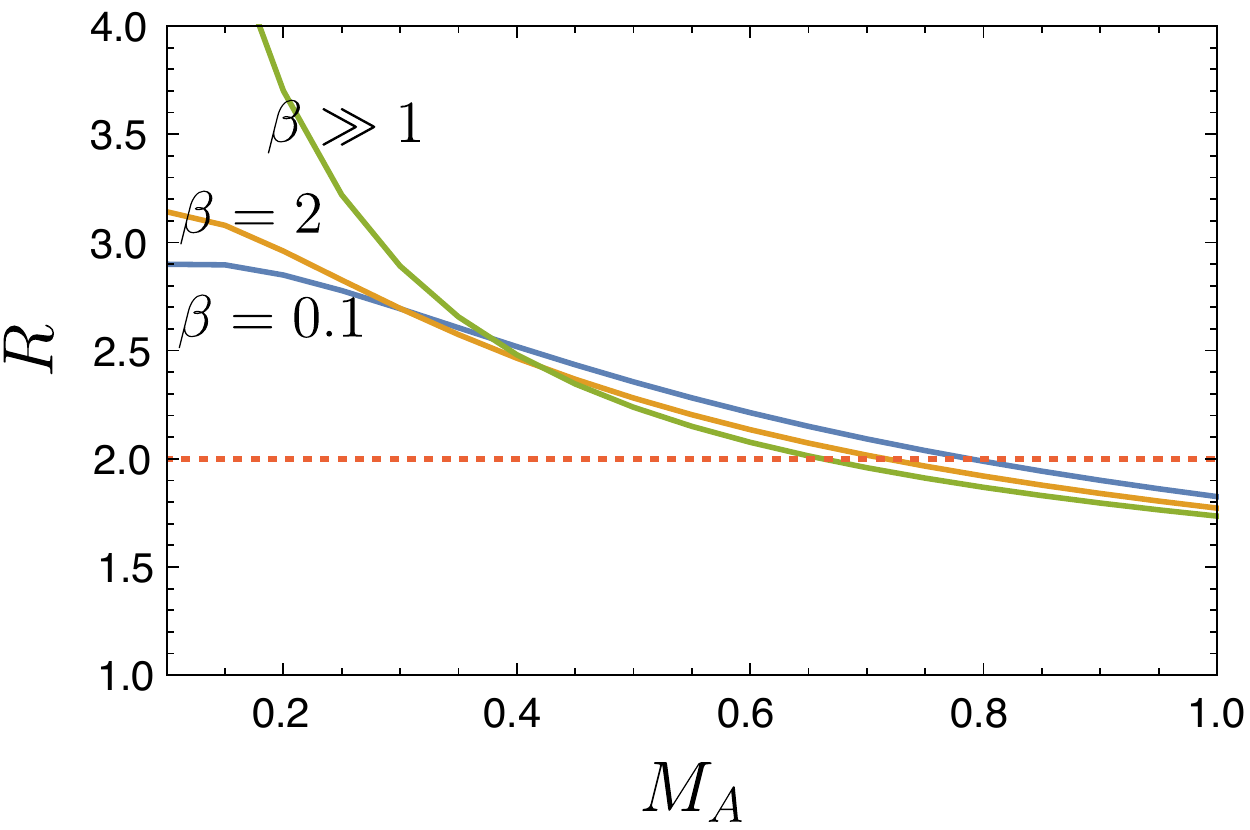}
\caption{Left: ratio of $E$ to $B$ power for equal mix of three MHD modes at various $\beta$ assuming uncorrelated density and magnetic field. Center: the same for equal mix of all three modes using the density model prescribed by \citetalias{caldwell2017dust}. The difference between our prescription and that of \citetalias{caldwell2017dust} is seen mostly at low $M_A$ for $\beta\sim 1$. Right: the same but assuming negligible density fluctuations. All figures are produced for $\gamma=-2$.}
\label{machplots}
\end{figure*}

In Fig. \ref{threemachplots}, we calculate contributions to $E$ and $B$ power, and their ratios, for each MHD mode individually at different plasma $\beta$ using power spectrum of velocity field given by equations \eqref{machplots}) and \eqref{anisgs}, and density fluctuation magnitude given by equation \eqref{deltanr}, but assuming uncorrelated density and magnetic field so that power in density and magnetic field add in square. These contributions are sensitive to $\beta$ only for slow and fast modes. First, since fast modes have isotropic power spectrum, the ratio they give is a flat number for a fixed $\beta$. At small $\beta$, $E$ and $B$ powers are mostly dominated by slow modes, for $\beta\sim 1$ by fast modes, and for large $\beta$ by Alfv\'en modes, except at low $M_A$, where fast modes contribute significantly to $B$ power. At low $\beta$, the E/B power ratio that this mode gives is $\sim 1.8$. Next, slow modes give rise to a ratio of $\sim 2$ or larger at $M_A<0.4$ at up to $\beta\sim 1$. Similarly, Alfv\'en mode gives a ratio of 2 for $M_A<0.5$. 

Exact mode composition of turbulence is another unknown. We first consider a possible scenario of equal mix of all three modes as shown in Fig. \ref{machplots}, where it is shown that the $E$ to $B$ ratio is $\gtrsim 2$ at $M_A\lesssim 0.5$. It is expected that the power scaling $P_f(k)\sim k^{-7/2}$ for fast modes is different from $P_a(k), P_s(k)\sim k^{-11/3}$ for Alfv\'en and slow modes \citep{cho2002compressible}. In this case, E/B power ratio will have weak scale dependence, which we neglect at the level of current discussion. Comparing the left-hand panel and centre panel of Fig. \ref{machplots}, we see that correlation of magnetic field and density has low impact in the E/B power ratio. However, within our model, there is a range of $M_A$ which fits with observed E/B power ratio even for $\beta\sim 1$, while the one adopted by \citetalias{caldwell2017dust} cannot reach this ratio for $\beta\sim 1$.

\begin{figure}
\centering
\includegraphics[scale=0.5]{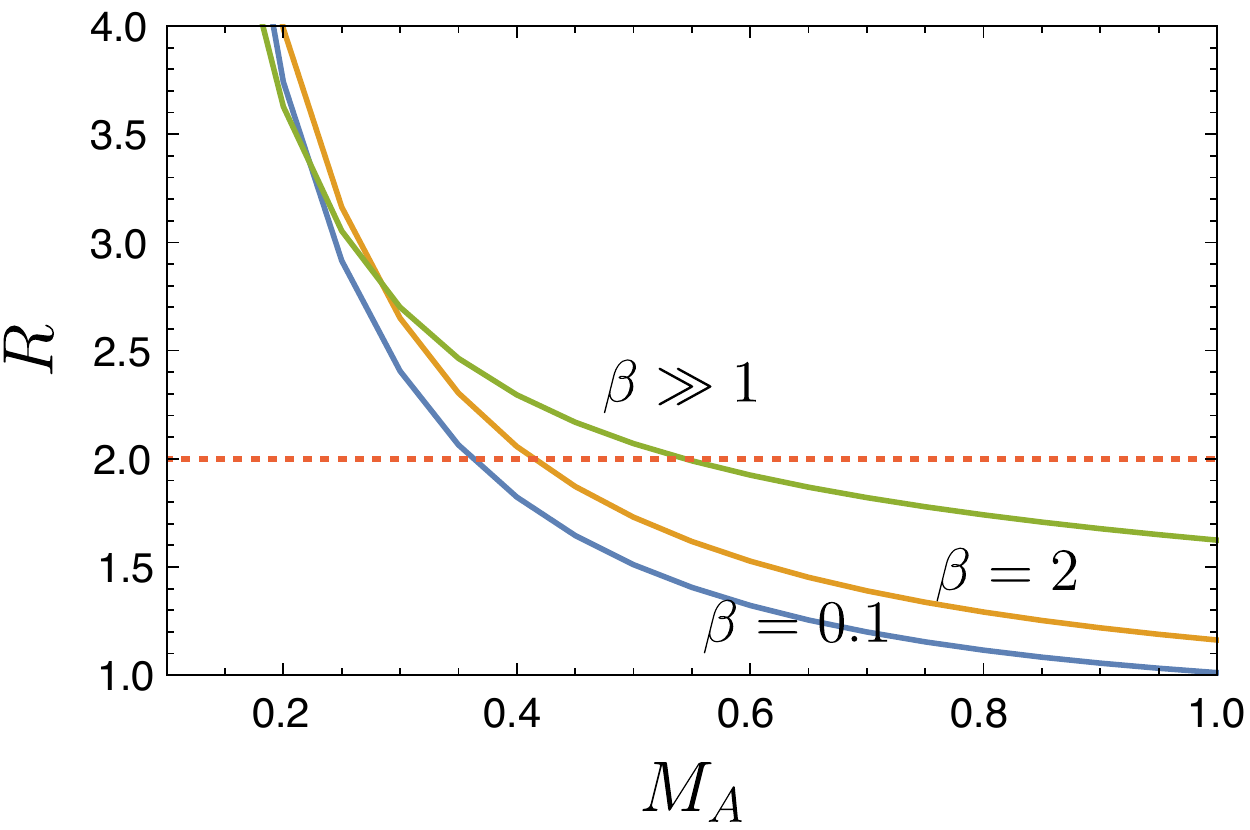}
\caption{E to B power ratio for equal mix of only Alfv\'en and slow modes at various $\beta$, for $\gamma=-2$. The dotted line represents the observed ratio of 2.}
\label{fig:alfvenslow}
\end{figure}

We also consider another possible scenario where isotropic fast modes are absent, and so that turbulence is mostly an equal mix of Alfv\'en and fast modes. As shown in Fig. \ref{fig:alfvenslow}, there is a wider range of $M_A$ for which ratio of 2 or more can be achieved. However, comparing Fig. \ref{fig:alfvenslow} and left-hand panel of Fig. \ref{machplots}, one can see that presence of fast modes drives the ratio closer to 2 at low $M_A$. 

Finally, we further test the effects of density fluctuations by considering limiting cases when dust density inhomogeneities are either dominant or
negligible relative to magnetic field inhomogeneities\footnote{In fact, dust density fluctuations could be induced by a completely independent process, and therefore, both amplitude and level of anisotropy of density field could be very different from what was modelled in Sec. \ref{mhdmodes}.}.  In the former case, if the density field is isotropic, the $E$ and $B$ power ratio will be driven towards unity, instead of 2. Maintaining a ratio of 2 will, thus, require fairly anisotropic dust density. For the latter case of negligible density fluctuations, we plot the E/B power ratio in the right-hand panel of Fig. \ref{machplots}. As shown in Fig. \ref{machplots}, the ratio of 2 is achieved for wider range of $M_A$, perhaps even for $M_A\sim 0.8$.

\section{Discussion and Summary}\label{sec:disc}
Studies of the Alfv\'en Mach number are still at its infancy. A number of promising ways to study $M_A$ have been suggested (see \citealt{esquivel2005velocity, esquivel2010tsallis, esquivel2011velocity}; \citealt{burkhart2014measuring}; \citealt{esquivel2015studying}; \citealt{lazarian2017synchrotron}) and tested with numerical simulations. However, we are not aware of the practical observational studies of $M_A$ using these techniques. Therefore, we have to rely on more indirect ways of evaluating the Alfv\'en Mach number. Observations show that while the turbulence in the disc, where driving takes place, is trans-Alfv\'enic, the magnetic field structure is less perturbed in the intermediate to high latitude (see \citealt{2016A&ARv..24....4B}). Thus, one expects to have sub-Alfv\'enic turbulence at high latitudes. More arguments in favour of sub-Alfv\'enic turbulence are provided in \citet{lazarian2016damping}, where it is shown that it is not possible to explain the data on cosmic ray isotropy unless the turbulence is sub-Alfv\'enic at high galactic latitudes. We accept that it is the direct measurements of $M_A$ that provides the acid test for our present explanation, and expect that our work will stimulate observational studies of $M_A$. A more detailed analysis of turbulence using polarization of dust as well as synchrotron polarization will be presented in our forthcoming paper.

In this Letter, using a realistic model of anisotropic velocity power spectrum in terms of physical parameter $M_A$, we have shown that for $M_A = 0.5$, the E/B power ratio can be about 2. As we argued above, this value of $M_A$ is reasonable for high galactic latitudes. Thus, we stress that the E/B power ratio results from Planck observation are consistent with the present day model of MHD turbulence, as long as the turbulence in the high latitude is sub-Alfv\'enic. Future observations shall provide us more information about the Alfv\'en Mach number at various Galactic latitudes. 

\footnotesize{
\nocite{*}
\bibliographystyle{mnras}
\bibliography{polarizationdraft}

\bsp	
\label{lastpage}

\end{document}